\documentclass{article}
\pdfoutput=1
\usepackage{spconf,amsmath}
\usepackage{graphicx}
\usepackage{amssymb}
\usepackage{tikz}	    % for drawings (similar to pstricks and Co.)
\usepackage{pgfplots}	% for drawings (similar to pstricks and Co.)
\usepackage{pgfkeys}	% for drawings (similar to pstricks and Co.)
\usepackage{import}
\usepackage{mathtools}
\usepackage[normalem]{ulem} % striking through text horizontally (AC)
\usepackage{cite}           % multiple citations more compact (AC)
\usepackage{siunitx}        % \SI environment
\usepackage[ruled,vlined]{algorithm2e}
\usepackage{multirow}

\def\IFFT{{\circ\hspace{-.33mm}\text{---}\hspace{-.33mm}\bullet}}

\title{CONTROL ARCHITECTURE OF THE DOUBLE-CROSS-CORRELATION PROCESSOR\\FOR SAMPLING-RATE-OFFSET ESTIMATION IN ACOUSTIC SENSOR NETWORKS}
\name{Aleksej Chinaev, Sven Wienand, and Gerald Enzner\thanks{Funded by German Research Foundation (DFG) - Project 282835863.}\vspace*{-1ex}}
\address{Ruhr-Universität Bochum, Department of Electrical Engineering and Information Technology,\\ 44801 Bochum, Germany, Email: \{aleksej.chinaev, sven.wienand, gerald.enzner\}@ruhr-uni-bochum.de\vspace*{-1ex}}

\newcommand\copyrighttext{%
  \footnotesize \copyright 2021\ IEEE. Personal use of this material is permitted. Permission from IEEE must be obtained for all other uses, in any current or future media, including reprinting/republishing this material for advertising or promotional purposes, creating new collective works, for resale or redistribution to servers or lists, or reuse of any copyrighted component of this work in other works.}
   
\newcommand\copyrightnoticee{%
\begin{tikzpicture}[remember picture,overlay]
\node[anchor=south, yshift=15pt] at (current page.south) {\mbox{\parbox{\dimexpr\textwidth-\fboxsep-\fboxrule\relax}{\copyrighttext}}};
\end{tikzpicture}%
}

\begin{document}
\ninept  % eingeschaltet, wie fast immer (Enzner)
\maketitle
\copyrightnoticee
\begin{abstract}
Distributed hardware of acoustic sensor networks bears inconsistency of local sampling frequencies, which is detrimental to signal processing. Fundamentally, sampling rate offset (SRO) nonlinearly relates the discrete-time signals acquired by different sensor nodes. As such, retrieval of SRO from the available signals requires nonlinear estimation, like double-cross-correlation processing (DXCP), and frequently results in biased estimation. SRO compensation by asynchronous sampling rate conversion (ASRC) on the signals then leaves an unacceptable residual. As a remedy to this problem, multi-stage procedures have been devised to diminish the SRO residual with multiple iterations of SRO estimation and ASRC over the entire signal. This paper converts the mechanism of offline multi-stage processing into a continuous feedback-control loop comprising a controlled ASRC unit followed by an online implementation of DXCP-based SRO estimation. To support the design of an optimum internal model control unit for this closed-loop system, the paper deploys an analytical dynamical model of the proposed online DXCP. The resulting control architecture then merely applies a single treatment of each signal frame, while efficiently diminishing SRO bias with time. Evaluations with both speech and Gaussian input demonstrate that the high accuracy of multi-stage processing is maintained at the low complexity of single-stage (open-loop) processing.
\end{abstract}

\section{Introduction and Relation to Prior Work}
\label{sec:intro}
\vspace*{-1ex}

Wireless acoustic sensor networks (WASNs) create new challenges in terms of synchronization between signals of different sensor nodes, but will exhibit advantages compared to concentrated microphone arrays \cite{Chen2003,Bertrand2011,Bertrand2015}. Specifically, WASNs pertain to different areas of application like source localization \cite{Griffin2015} and separation \cite{Wehr2004}, speech enhancement \cite{Tavakoli2016} and recognition \cite{Araki2017}, blind channel identification \cite{ThuEnz2018}, or scene classification \cite{Gergen2015}. Since all sensors in a WASN obey to their own clocks with slightly imprecise oscillators, sampling rate offsets (SRO) and accumulated sampling time offsets (STO) are inevitable. The sensor signals thus require synchronization before successful utilization in the respective application \cite{Lienhart2003,Zeng2015,Schmalenstroeer2018}.

Generally, two groups of synchronization architectures can be distinguished, the first of which is described as open-loop (OL) synchronization. Here, SRO and STO are blindly estimated from just the available acoustic signals by different algorithms like coherence drift \cite{MarkovichGolan2012, Bahari2017, Schmalenstroeer2017}, maximum-likelihood \cite{Miyabe2015, Araki2019}, correlation-based \cite{Cherkassky2014, Wang2016b, Cherkassky2017}, or double-cross-correlation processing (DXCP) \cite{Chinaev19}. Afterwards, raw signals are resampled by ASRC, e.g., \cite{Oppenheim1999,ChiThuEnz2018,ChiEnzSch2018}.

The second group uses a closed-loop (CL) architecture. Here, the resampled sensor signal feeds into the SRO/STO estimation to enhance its accuracy and therefore getting a better synchronization. More specifically, a so-called multi-stage (i.e., offline CL) procedure estimates SRO/STO across the whole sensor signal and resamples it and reiterates, which rests upon the observation that SRO estimation bias vanishes with smaller input SRO \cite{Miyabe2015, Wang2016b}. Any OL estimator of SRO/STO can be embedded into a multi-stage (MS) form as shown in \cite{Schmalenstroeer2017} for the estimator in \cite{MarkovichGolan2012}.

In contrast, research in \cite{Cherkassky2017} and \cite{Schmalenstroeer2015} applies sample- or frame-wise estimation of SRO/STO to resample the sensor signals and continuously update the estimation based on resampled signals. This results in an online CL synchronization. The systems are, however, different in that \cite{Cherkassky2017} relies on centralized processing of acoustic waveforms, while \cite{Schmalenstroeer2015} pursues decentralized SRO estimation in the context of two-way message passing of timestamps.

In this paper, we focus on SRO between signals for its unique source in clock deficiency, while STO often exhibits many other causes, such as network delays. The proposed blind synchronization scheme can be classified as online closed-loop and for this purpose introduces an online form of the recently deployed DXCP method \cite{Chinaev19} for SRO estimation. The framewise updated SRO estimate then instantly controls the preceding resampling module (or potentially a controllable analog-to-digital converter) of the closed-loop system and thus is able to cope with time-varying SRO. To link the asynchronous resampling with the SRO estimation module, this paper derives a dynamical model of DXCP and a respective SRO controller based on internal model control (IMC) theory \cite{Francis1976,Morari1989}. 

The paper is organized as follows: Sec.\ \ref{sec:preliminaries} introduces the online form of DXCP in open and closed loop with resampling. Our dynamical model of DXCP is derived in Sec.\ \ref{sec:model} and the related IMC controller described in Sec.~\ref{sec:controller}. Performance of the resulting online closed-loop system is then evaluated in Sec.\ \ref{sec:evaluation} with respect to (online) open-loop and offline MS forms, before Sec.\ \ref{sec:conclusion} concludes.
\vspace*{-1ex}
\section{Signal and Control System Models}
\label{sec:preliminaries}
\vspace*{-1ex}

To prepare our design of the IMC controller, this section initially relies on a discrete-time audio system model and eventually converts to a frame-wise system model of the underlying SRO process, while the online form of DXCP method is introduced for SRO assessment.\vspace*{-1ex}

\subsection{Online form of DXCP in an open-loop architecture}
\label{subsec:OnlineDXCP}

An open-loop synchronization architecture using DXCP is depicted in Fig.~\ref{fig:OL_DXCP}~a). The signal $z_1(n)$ at sampling-time $n$ is assumed to be a reference signal obtained via analog-digital conversion (ADC) at the nominal sampling-time interval $T_1 \!=\! 1/f_s$. The sequence $z_2(n)$ is delivered by an imperfect ADC with slightly different sampling-time interval $T_2 = (1 + \varepsilon) \cdot T_1$, where the deviation $\varepsilon \not= 0$ is termed the SRO, since the parameter $\varepsilon$ similarly relates sampling frequencies as $f_\text{s,2} \!=\! 1/T_2 \!=\! (1 \!-\! \varepsilon/(1 \!+\! \varepsilon)) \!\cdot\! f_\text{s} \!\approx\! (1 \!-\! \varepsilon) \!\cdot\! f_\text{s}$ due to $|\varepsilon| \!\ll\! 1$. Further processing and modeling takes place on a frame basis with $N_\text{s}$ samples frame-shift, thus time interval $T_\text{A}\!=\!T_1 \!\cdot\! N_\text{s}$. On this larger time-scale, SRO induces an accumulating time drift (ATD) 
\begin{align}
    \tau(\ell)
    %= \varepsilon \cdot \left( \frac{N-1}{2} + N_\text{S} \cdot (\ell-1) \right)
    %= \varepsilon \cdot \left[ (N \! - \! 1)/2 + N_\text{S} \cdot (\ell \! - \! 1) \right]
    = \tau(\ell \! - \! 1) + N_\text{s} \cdot \varepsilon
    \label{eq:tau_ell}
\end{align}
between the $\ell$-th frames of $z_1(n)$ and $z_2(n)$. Since a difference of consecutive ATDs $\tau(\ell) - \tau(\ell-1)$ then obviously provides a good handle to fixed SRO~$\varepsilon$, we derived in \cite{Chinaev19} a five-step time-domain DXCP for SRO retrieval using the entire signals $z_1(n)$ and $z_2(n)$, which is however obstructive for online processing. To overcome this issue, we shall here translate the first four DXCP steps into the short-time Fourier transform (STFT) domain and deliver a frame-wise SRO estimation $\widehat{\varepsilon}(\ell)$ in step 5 (returned to time domain):

\begin{figure}[t!]
\begin{minipage}[b]{1.0\linewidth}
  \begin{center}
    (a)\medskip\vspace{-2.5mm}
    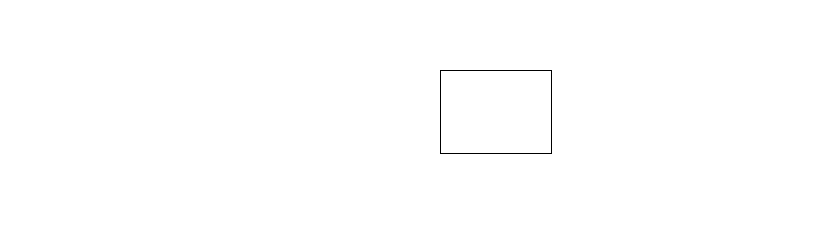
    %\psfragfig[width=\columnwidth]{./pics/1a_DXCP_OL_TEX}{}
  \end{center}
\end{minipage}
\begin{minipage}[b]{1.0\linewidth}
  \begin{center}
    (b)\medskip\vspace{-1mm}
    % Pfad auf CD: /Matlab/work/Estimation/Paper_Fig1b.m
    \input{./pics/1b_DXCP_OL_MultiStepV2.tikz}
  \end{center}
\end{minipage}
\vspace*{-3ex}
\caption{(a) Online DXCP in open loop, (b) Dynamic response to SRO $\varepsilon \in \{0, 5, 10, 20, 40\}\,\text{ppm}$ for white noise acoustic point source.}
\vspace*{-1ex}
\label{fig:OL_DXCP}
\end{figure}

\textbf{Step~1}: Compute frame-wise $N$-point STFT coefficients $Z_1(k,\ell)$ and $Z_2(k,\ell)$ at discrete frequency $k$ from $z_1(n)$ and $z_2(n)$.

\textbf{Step~2}: Calculate a short-time, primary cross spectral density~(CSD) via recursive averaging with smoothing constant $\alpha_1$, 
\begin{align}
    \widetilde{\Phi}_{12}&(k,\ell)
    %= \alpha_1 \hspace{-.5mm} \cdot \hspace{-.5mm} \widetilde{\Phi}_{12}(k,\ell \hspace{-.5mm} - \hspace{-.5mm} 1) \hspace{-.5mm} + \hspace{-.5mm} (1 \hspace{-.5mm} - \hspace{-.5mm} \alpha_1) \hspace{-.5mm} \cdot \hspace{-.5mm} Z_1(k,\ell) \hspace{-.5mm} \cdot \hspace{-.5mm} Z_2^*(k,\ell) \;,
    %
    = \alpha_1 \! \cdot \! \widetilde{\Phi}_{12}(k,\ell \! - \! 1) \! + \! (1 \! - \! \alpha_1) \! \cdot \! Z_1(k,\ell) \! \cdot \! Z_2^*(k,\ell) \;,
    \label{eq:1CSD}
\end{align}
where $()^*$ denotes complex conjugation and $\widetilde{\Phi}_{12}(k,0) = 0$. The maximum position $\widehat{\tau}$ of the corresponding correlation function in the time domain would be a handle to the current time drift $\tau(\ell)$.

\textbf{Step~3}: Save the $L_\text{b}+1$ latest, complex-valued CSDs $\widetilde{\Phi}_{12}(k,\ell')$ for $\ell' \in \{\ell-L_\text{b}, \ldots , \ell \}$ in a circular buffer for further processing.

\textbf{Step~4}: Obtain a secondary CSD $\widetilde{\Psi}_{12}(k,\ell)$ via recursive averaging of the product of two primary CSDs at a distance of $L_\text{b}$ frames,
\begin{align}
      \widetilde{\Psi}_{12}(k,\ell) = \alpha_2 &\cdot \widetilde{\Psi}_{12}(k,\ell-1) +
      \label{eq:2CSD}\\
      +&(1-\alpha_2) \cdot \widetilde{\Phi}_{12}(k,\ell) \cdot \widetilde{\Phi}_{12}^{*}(k,\ell-L_\text{b}) \;,
      \nonumber
\end{align}
with smoothing constant $\alpha_2$ and $\widetilde{\Psi}_{12}(k,L_\text{b}) = 0$. A corresponding correlation function $\widetilde{\psi}_{12}(\ell,\lambda)$ at lag $\lambda$ is then computed per frame $\ell$ as the $N$-point inverse fast Fourier transform (IFFT) of \eqref{eq:2CSD}, i.e.,
\begin{align}
      \widetilde{\psi}_{12}(\ell,\lambda) &\stackrel{\text{IFFT}}{\IFFT}\widetilde{\Psi}_{12}(k,\ell) \;.
\label{eq:IFFT2CCF}
\end{align}
Considering \eqref{eq:tau_ell} and \eqref{eq:2CSD}, the maximum position of $\widetilde{\psi}_{12}(\ell,\lambda)$ then amounts to a handle to the ATD differentiator
\begin{align}
    \tau_\Delta(\ell)
    = \tau(\ell) - \tau(\ell-L_\text{b})
    = L_\text{b} \cdot N_\text{s} \cdot \varepsilon \;.
    \label{eq:tau_Delta}
\end{align}

\textbf{Step~5}: Applying a parabolic maximum search as in \cite{Chinaev19} over $\widetilde{\psi}_{12}(\ell,\lambda)$, the real-valued $\tau_\Delta(\ell)$ is estimated for $\ell \geq L_\text{b}\!+\!L_\text{c}\!+\!1$ and from \eqref{eq:tau_Delta} we deduce the related SRO estimate in the steady state as
\begin{align}
    \widehat{\varepsilon}(\ell)
    = \frac{ \widehat{\tau}_\Delta(\ell) }{N_\text{s}\cdot L_\text{b}}\;.
    \label{eq:SRO_ell}
\end{align}

The system in Fig.~\ref{fig:OL_DXCP}~a) eventually uses the estimated SRO for continuous feedforward control of the resampler to deliver a signal $z_{2,\text{s}}(n)$ synchronized to the reference $z_1(n)$ by ASRC \cite{Oppenheim1999}.  

An experiment in Fig.~\ref{fig:OL_DXCP}~b) confirms that online DXCP\footnote{For $f_\text{s} \!=\! \SI{16}{\text{kHz}}$, the DXCP parameters are set to $N\!=\!2^{13}$, $N_\text{s}\!=\!2^{11}$, $L_\text{b}\!=\!39$, $L_\text{c}\!=\!19$, $\alpha_1\!=\!0.5$ and $\alpha_2\!=\!0.99$.} is able to estimate actual SRO in a wide range on the relevant parts per million (ppm) scale. In the steady state, an estimation bias in the order of $\pm\SI{1}{\text{ppm}}$ can be seen for SRO $\varepsilon \not= \SI{0}{\text{ppm}}$. That bias diminishes for actual SRO $\varepsilon = \SI{0}{\text{ppm}}$, which motivates a control architecture with resampling in the loop. The time evolution in Fig.~\ref{fig:OL_DXCP}~b) further reveals an SRO-dependent, thus nonlinear dynamical behaviour of online DXCP, for instance, almost exponential response for small and almost dead-time characteristic for large actual SRO, respectively. The latter phenomena occurs because the maximum search of step $5$ of the DXCP, after large SRO step change, can switch to a new global maximum of $\widetilde{\psi}_{12}(\ell,\lambda)$ only towards the new steady state.

\begin{figure}[t!]
\begin{minipage}[b]{1.0\linewidth}
  \begin{center}
    (a)\medskip\vspace{-2.1mm}
    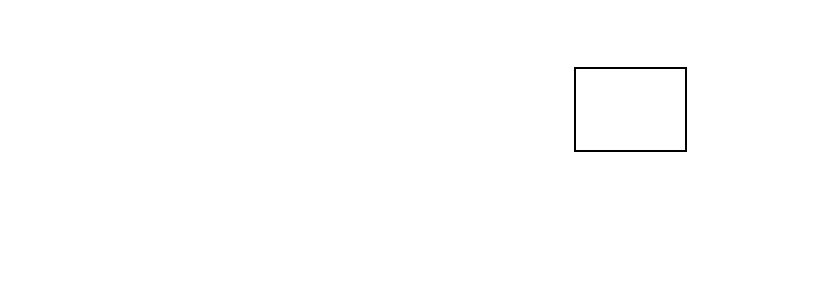
  \end{center}
\end{minipage}
\begin{minipage}[b]{1.0\linewidth}
  \begin{center}
    (b)\medskip\vspace{-2.1mm}
    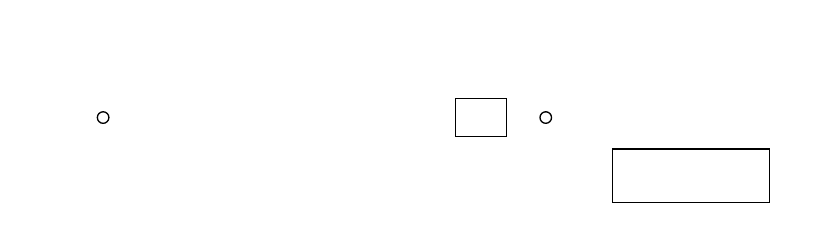
  \end{center}
\end{minipage}
\caption{(a) DXCP in a closed-loop form, (b) Block diagram of closed-loop DXCP architecture in the domain of control signals.}
\vspace*{-1ex}
\label{fig:CL_DXCP}
\end{figure}

\subsection{Online closed-loop architecture with DXCP method}

In a closed-loop control architecture as shown in Fig.~\ref{fig:CL_DXCP}~a), the signal $z_2(n)$ gets resampled first. The resulting $z_{2,\text{s}}(n)$ is then fed to the DXCP, which now measures the residual SRO $\Delta\varepsilon(\ell)$ between $z_1(n)$ and $z_{2,\text{s}}(n)$. The estimate $\Delta\widehat{\varepsilon}(\ell)$ supports the controller, which in turn delivers the control signal $\widehat{\varepsilon}(\ell)$ for resampling. In the steady state, the system is meant to approach $\Delta\widehat{\varepsilon}(\ell)\to 0$ and $\widehat{\varepsilon}(\ell) \to\varepsilon$.

Abstracting the underlying SRO from the source audio signals, we can create a block diagram of the control loop of Fig.~\ref{fig:CL_DXCP}~a) in the domain of control signals, see Fig.~\ref{fig:CL_DXCP}~b). Here, the function of the resampler can be effectively described as a subtraction of the estimated SRO~$\widehat{\varepsilon}(\ell)$ from the actual SRO~$\varepsilon(\ell) = \varepsilon$. As a synchronisation of $z_1(n)$ and $z_{2,\text{s}}(n)$ supports the residual SRO $\Delta \widehat{\varepsilon}(\ell)$ to approach zero, the desired signal $w(\ell)$ is here defined as zero.
\section{Linear dynamic model of online DXCP}
\label{sec:model}
\vspace*{-0.5ex}

In order to design in Sec.~\ref{sec:controller} an IMC controller for the system in Fig.\ref{fig:CL_DXCP}, a linear DXCP model in the domain of control signals is required. Such model can be derived for small SRO with Eqs.\ \eqref{eq:tau_ell} to \eqref{eq:SRO_ell}. 

Considering the signal $\varepsilon \!=\! \varepsilon(\ell)$, the integration process \eqref{eq:tau_ell} and the additional first-order recursive average \eqref{eq:1CSD} can be together represented as a $z$-domain transfer function (TF) from $\varepsilon$ to $\widehat{\tau}$,
\begin{align}
    G_{\widehat{\tau},\varepsilon}(z)
    = N_\text{s} \cdot \frac{z}{z-1}\cdot \frac{(1-\alpha_1) \cdot z}{z-\alpha_1}\;,
    \label{eq:G1_DXCP}
\end{align}
which is based on the idea that $\widetilde{\Phi}_{12}(k,\ell)$ captures the SRO-induced ATD $\tau(\ell)$. A further TF from $\widehat{\tau}$ to $\widehat{\varepsilon}$ is obtained with \eqref{eq:tau_Delta}, \eqref{eq:2CSD} and \eqref{eq:SRO_ell},
\begin{align}
	 G_{\widehat{\varepsilon},\widehat{\tau}}(z)
	 = \frac{1}{N_\text{s} \cdot L_\text{b}} \cdot \frac{z^{L_\text{b}}-1}{z^{L_\text{b}}} \cdot \frac{(1-\alpha_2)\cdot z}{z-\alpha_2}\;,
	 \label{eq:G2_DXCP}
\end{align}
and we eventually form a product of these individual TFs \eqref{eq:G1_DXCP} and \eqref{eq:G2_DXCP} to arrive at the sought linear dynamical model $G_\text{DXCP}(z)$:
\begin{align}
    G_{\text{DXCP}}(z)
    = \frac{1}{L_\text{b}}
    \hspace{-.5mm}\cdot\hspace{-.5mm} \frac{z}{z-1}
    \hspace{-.5mm}\cdot\hspace{-.5mm} \frac{(1-\alpha_1) \hspace{-.5mm}\cdot\hspace{-.5mm} z}{z-\alpha_1}
    \hspace{-.5mm}\cdot\hspace{-.5mm} \frac{z^{L_\text{b}} \hspace{-.5mm} - \hspace{-.5mm} 1}{z^{L_\text{b}}}
    \hspace{-.5mm}\cdot\hspace{-.5mm} \frac{(1-\alpha_2) \hspace{-.5mm}\cdot\hspace{-.5mm} z}{z-\alpha_2}\;.
    \label{eq:G_DXCP}
\end{align}

The entire model bears several time-constants related to $\alpha_1$, to $L_\text{b}$ (resulting jointly from integrator and differentiator) and to $\alpha_2$. In order to reduce the model order and the complexity of the IMC controller to a minimum, we here suggest a model approximation
\begin{align}
    \widehat{G}_\text{DXCP}(z) = \frac{1-\alpha_2}{z-\alpha_2}\;,
    \label{eq:G_DXCP_hat}
\end{align}
which is determined by the dominant time-constant of $G_\text{DXCP}(z)$.
\vspace*{-0.5ex}
\section{Controller design}
\vspace*{-0.5ex}
\label{sec:controller}

We first deliver the fundamental design of an IMC controller for the closed-loop control architecture in Fig.~\ref{fig:CL_DXCP} and then argue some of its parameters based on robustness considerations. \vspace*{-1ex}

\subsection{Linear IMC feedback control}

\begin{figure}[t!]
  \begin{center}
    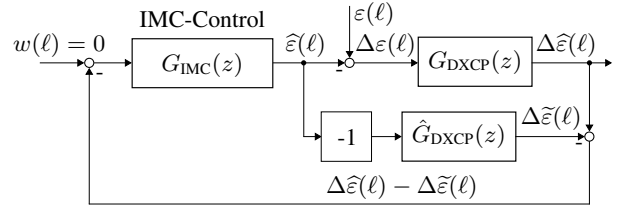
  \end{center}
  \vspace{-3mm}
  \caption{Online DXCP in a IMC control-loop architecture.}
  \label{fig:DXCP_IMC}
\end{figure}

Returning our mind to the overall nonlinear behavior of DXCP as illustrated by Fig.~\ref{fig:OL_DXCP}~b), the algorithmic models in \eqref{eq:G_DXCP} or \eqref{eq:G_DXCP_hat} are both meant to represent merely the linear working range of DXCP (i.e., small SRO changes under $1/(N_\text{s}\cdot L_\text{b})\!=\!12.5\,$ppm with $L_\text{b}\!=\!39$, $N_\text{s}\!=\!2^{11}$) for continuous control purpose. For this range, we here foresee the design of a control law according to IMC theory \cite{Francis1976,Morari1989} with a circuit shown in Fig.~\ref{fig:DXCP_IMC}. It implies a plant predictive $\widehat{G}_\text{DXCP}(z)$ leg in parallel to the actual DXCP and the output error feeds back to an IMC filter $G_\text{IMC}(z)$ to be designed for quadratic minimization of the control error, i.e., the residual SRO signal $\Delta\varepsilon(\ell)\!=\!\varepsilon(\ell)\!-\!\widehat{\varepsilon}(\ell)$. The signal $\Delta\varepsilon(\ell)$ can be calculated in closed-form in $z$-domain as
\begin{align}
    \Delta E(z)
    %= \frac{1+G_\text{IMC}(z) \cdot \widehat{G}_\text{DXCP}(z)}{1 + G_\text{IMC}(z) \cdot \widehat{G}_\text{DXCP}(z) - G_\text{IMC}(z) \cdot G_\text{DXCP}(z)} \cdot E(z).
    = \frac{1+G_\text{IMC}(z) \cdot \widehat{G}_\text{DXCP}(z)}{1 + G_\text{IMC}(z) \cdot (\widehat{G}_\text{DXCP}(z) - G_\text{DXCP}(z))} \cdot E(z)
    \label{eq:Ez_to_DelEz}
\end{align}
or, assuming ideal prediction with $\widehat{G}_\text{DXCP}(z) \!=\! G_\text{DXCP}(z)$,
even as
\begin{align}
    \Delta E(z)=(1+G_\text{IMC}(z) \cdot G_\text{DXCP}(z))\cdot E(z) \;.
    \label{eq:deltaE_z_assumption}
\end{align}

The latter yields a very simple solution $G_\text{IMC}^\text{opt}(z) \!=\! -G_\text{DXCP}^{-1}(z)$, which is independent of the actual SRO $E(z)$. In order to deal appropriately with feasibility of the control circuit, the optimal solution can be extended with an additional PT$_{n_f}$ filter function
\begin{align}
    F(z) = \left( \frac{1 - \beta}{z - \beta} \right)^{\hspace{-1mm}n_\text{f}} \hspace{-2mm}\;,
    \label{eq:AddFilter}
\end{align}
with constant $\beta\!=\!\mathrm{e}^{-T_\text{A}/T_\text{f}}$, $n_\text{f}$ the filter order, and $T_\text{f}$ the desired filter time-constant on the continuous time scale. Thus,
\begin{align}
    G_\text{IMC}(z)
    = G_\text{IMC}^\text{opt}(z) \cdot F(z)
    = -\frac{F(z)}{G_\text{DXCP}(z)}\;.
    \label{eq:controller_assumption}
\end{align}
Since the closed-loop TF in \eqref{eq:deltaE_z_assumption} in this case amounts to $1\!-\!F(z)$, the time-constant $T_\text{f}$ defines the speed of the entire IMC controller.

\subsection{Robustness against model uncertainty}
\label{subsec:RobustIMC}

Related to the former arguments of complexity and feasibility of the closed-loop control architecture, we do now revert to the model approximation \eqref{eq:G_DXCP_hat} and practically calculate our IMC controller as
\begin{align}
    G_\text{IMC}(z)
    %= -\frac{F(z)}{G_\text{DXCP}(z)}
    = -\frac{F(z)}{\widehat{G}_\text{DXCP}(z)}\;.
    \label{eq:approx_controller}
\end{align}

To assess robustness of the control loop under the approximation, we can  express a model error $\delta G_M(z)$ in multiplicative form
\begin{align}
    G_\text{DXCP}(z) = \widehat{G}_\text{DXCP}(z)\cdot(1+\delta G_M(z))
    \label{eq:model_uncertainty}
\end{align}
and determine $\delta G_M(z)$ with \eqref{eq:G_DXCP} and \eqref{eq:G_DXCP_hat}.
Relying on this multiplicative model uncertainty, the authors in \cite{Morari1989} showed that stability of the IMC control-loop is met when
\begin{align}
    |F(z)| \cdot |\delta G_M(z)| < 1 \text{ for } |z| = 1\;.
\end{align}

In order to take both the linear and nonlinear model approximations of DXCP into account, we here require the critical product $|F(z)| \cdot |\delta G_M(z)|$ to be much smaller than $1$, in the interest of higher robustness. It turns out that the overall behavior of the closed-loop control architecture is determined mainly by the parameters $n_\text{f}$ and $T_\text{f}$ of filter \eqref{eq:AddFilter} and, therefore, is a trade off between stability and settling time of $F(z)$. Accordingly, we choose $n_\text{f} = 2$ and $T_\text{f} = \SI{8}{\second}$, which yields $\max_{|z|=1} ( |F(z)| \cdot |\delta G_M(z)| ) = 0.1514$ with moderate filter order and a relatively small settling time, sufficiently good for the application of long-term SRO control at hand.

\subsection{Entire closed-loop architecture with online DXCP}

\begin{figure}[t!]
  \begin{center}
    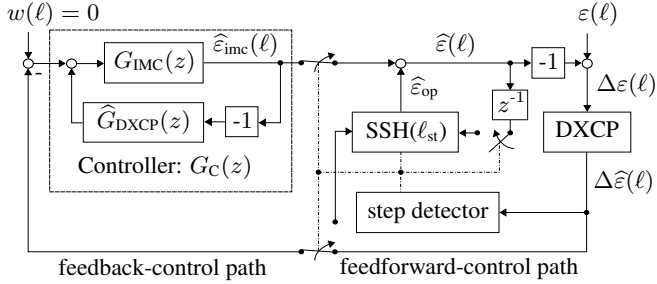
  \end{center}
  \vspace{-4mm}
  \caption{Online DXCP with IMC-based feedback-control and extension by a detection-based "large-SRO" feedforward-control path.}
  \label{fig:overall_system2}
\end{figure}

\begin{figure}[t!]
  \begin{center}
    % Pfad auf CD: /Matlab/work/Estimation/Paper_Fig5.m
    \input{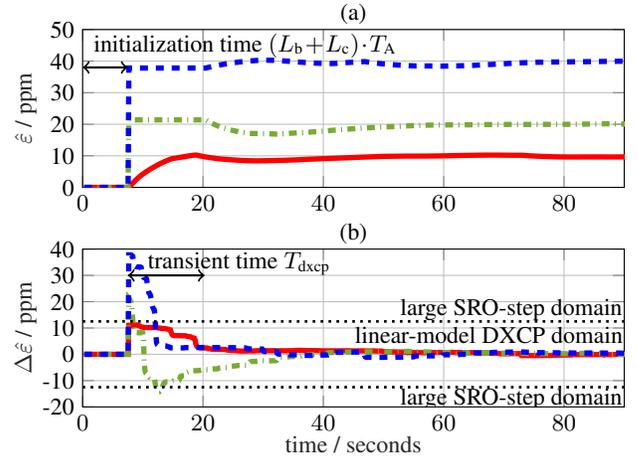}
  \end{center}
  \vspace{-4mm}
  \caption{Convergence of control signals (a) $\widehat{\varepsilon}(\ell)$ and (b) $\Delta\widehat{\varepsilon}(\ell)$ for SRO values $\varepsilon \in \{10, 20, 40\}\,\text{ppm}$ and for speech acoustic source.}
  \label{fig:CL_Convergence}
\end{figure}

In order to deal with the overall nonlinear dynamical characteristic of DXCP as shown by Fig.~\ref{fig:OL_DXCP}~b), we eventually propose with Fig.~\ref{fig:overall_system2} a hybrid system with two switched, yet interactive modes, specifically a linear IMC feedback-control subsystem $G_C(z)$ equivalent to Fig.~\ref{fig:DXCP_IMC} to address small SRO, and a feedforward-control subsystem to treat large SRO changes. For switching between the feedback- and forward-control modes, an SRO step detector invoked on the condition $|\Delta \widehat{\varepsilon}(\ell_\text{st})| \!>\! 1/(N_\text{s}\cdot L_\text{b})\!=\!12.5\,$ppm of the DXCP output with $L_\text{b}\!=\!39$, $N_\text{s}\!=\!2^{11}$, where $\ell_\text{st}$ then denotes the frame index of a positive detection. Because of the delicate dead-time characteristic of DXCP related to an observed SRO step-change, all switches turn over synchronously to detach the IMC feedback-control path for a duration of the time-constant $T_\text{dxcp} \!=\! T_\text{A}/\text{ln}(1/\alpha_2)\!=\!\SI{12.73}{\second}$ of DXCP with $\alpha_2 \!=\! 0.99$, $f_s\!=\!16\,$kHz. During this time, the detected SRO change $|\Delta \widehat{\varepsilon}(\ell_\text{st})|$ is sampled and summed to the latest actuation $\widehat{\varepsilon}(\ell)$ and held fixed as $\widehat{\varepsilon}_\text{op}$ by the sample-sum-and-hold (SSH) module. For the example of the system initialization, Fig.~\ref{fig:CL_Convergence}~a) illustrates the different behavior for small and large initial SRO. After the transient time $T_\text{dxcp}$ of DXCP output $\Delta \widehat{\varepsilon}(\ell)$, seen in Fig.~\ref{fig:CL_Convergence}~b), the IMC feedback control is reset to zero and activated by setting all switches back into their initial position. It starts continuous operation of its $\widehat{\varepsilon}_\text{imc}(\ell)$ output in addition to the fixed feedforward point of operation $\widehat{\varepsilon}_\text{op}$ in form the summing actuation $\widehat{\varepsilon}(\ell)$, as illustrated by Fig.~\ref{fig:CL_Convergence}, before update of $\widehat{\varepsilon}_\text{op}$ would take place with new step detection.\vspace*{-1ex}
\section{Experimental Evaluation}
\label{sec:evaluation}
\vspace*{-1ex}

For acoustic room setup in Figs.~\ref{fig:OL_DXCP}~a) and \ref{fig:CL_DXCP}~a),  experimental data is generated by simulation of a $(4 \! \times \! 5 \! \times \! 3)\,\text{m}$ room size with a reverberation time $T_{60} \hspace{-.5mm} = \hspace{-.5mm} \SI{200}{\text{ms}}$ using the image source method \cite{Allen1979}. In doing so, the source and sensor positions are chosen randomly, while for the acoustic point source signal we rely on three minutes of either computer generated white noise or TIMIT speech \cite{Zue1990}, respectively, at a reference sampling rate $f_\text{s} \!=\! \SI{16}{\text{kHz}}$. The convolved microphone signals are then superimposed with  spherical diffuse babble noise at $\text{SNR}_\text{diff} \!=\! \SI{15}{\text{dB}}$ according to \cite{Habets2008} and with uncorrelated sensor self-noise at $\text{SNR} \!=\! \SI{20}{\text{dB}}$. The signals of the second (i.e., asynchronous) node from Figs.~\ref{fig:OL_DXCP}~a) and \ref{fig:CL_DXCP}~a) are further manipulated with different SRO $\varepsilon \in \{ 0 \hspace{-.5mm} : \hspace{-.5mm} 20 \hspace{-.5mm} : \hspace{-.5mm} 100 \} \, \text{ppm}$ imposed by a Hann-windowed SINC interpolation~\cite{Oppenheim1999} of window length $N_{w,1} \!=\! 513$ samples.

For comparison with state-of-the-art, we consider both open- and closed-loop (including multi-stage) architectures, where three coherence-drift methods (CD) and former double-cross-correlation processing (DXCP) are evaluated: the least-squares CD (LCD) and weighted LCD (WLCD) \cite{Bahari2017}, the weighted average CD (WACD) \cite{Schmalenstroeer2017} and the time-domain offline DXCP \cite{Chinaev19}. While parameters of all CD-based methods are set as in \cite{Schmalenstroeer2017}, the offline DXCP is configured as in \cite{Chinaev19}. For open-loop CD-based methods, the current SRO estimation is based on an aggregate of the $17$ most recent coherence frames to roughly match the data-window, i.e., the time-constant $T_\text{dxcp}$ of the recursive averaging of the proposed online DXCP, at coherence-frame shift of $1.5$s. The closed-loop methods further apply a SINC resampler ($N_{w,2}\!=\!17$) inbetween the executions of SRO estimation. The MS approaches are each stopped after suitable convergence, i.e., MS-LCD after $15$, MS-WLCD after $10$, MS-WACD after $8$, and MS-DXCP after $3$ iterations. Parameters of the proposed system are set for online DXCP as in Fig.~\ref{fig:OL_DXCP}~b), while the IMC controller uses parameters from Sec.~\ref{subsec:RobustIMC}.

Tab.~\ref{tab:Evaluation} shows results in various dimensions, e.g., final root-mean-square error (RMSE) by SRO estimators (i.e., steady state or final MS iteration) over 60 random acoustic configurations. The "noise" column firstly confirms strong advantage of closed-loop over open-loop processing. The open-loop "speech" column shows enhanced accuracy of DXCP (online \& offline) over LCD at similar "realtime factors" (RTF, the lower the better). An "online" checkmark in parenthesis next to LCD refers to restricted online ability because of large coherence frame-shift. The closed-loop "speech" column shows a break even of proposed online DXCP- w.r.t. MS-processing in terms of RMSE, however, maintaining the low RTF of open-loop processing. The RTF column "w/o SINC" eventually refers to an interesting configuration with SRO-controllable hardware support for resampling in the loop, which is naturally not feasible with MS.

\vspace{-2mm}

\begin{table}[h!]
    \centering
    \begin{tabular}{c|c|c|c|c||c|c}
    & architecture    & \!\!online\!\!  & \multicolumn{2}{c||}{ \!RMSE\,/\,ppm\!}                   & \multicolumn{2}{c}{ RTF$\times  10^{-3}$}\\
    & and method      & \!\!ability\!\! &    \!noise   \!                    & \! speech\!\!   &   \!w/\,SINC\!    &    \!w/o\!  \\
    \hline
    \hspace{-1mm}\multirow{4}{*}{\rotatebox[origin=c]{90}{open-loop}}\hspace{-1mm}
    &LCD  \cite{Bahari2017}  & (\checkmark) &       \!0.86\!      &      \!19.86\!     &       \!129.7\!       &   \!5.3\!\\
    &WLCD \cite{Bahari2017}  & (\checkmark) &       \!1.41\!      &      \!17.50\!     &       \!129.7\!       &   \!5.4\!    \\
    &offl. DXCP \cite{Chinaev19}   & -      &  \!0.52\!      &       \!1.10\!     &       \!131.4\!       &   \!7.0\!    \\ \cline{2-7}
    &prop. DXCP              &  \checkmark  &       \!0.94\!      &       \!\bf 0.59\!     &       \!127.1\!       &   \!2.7\!     \\
    \hline
    \hspace{-1mm}\multirow{5}{*}{\rotatebox[origin=c]{90}{closed-loop}}\hspace{-1mm}
    &\!\!MS-LCD\cite{Bahari2017} \!\!                 &       -      &       \!$<0.1$\!      &       \!0.32\!     &    \!1936.8\!       &  -\\
    &\!\!MS-WLCD\cite{Bahari2017}  \!\!               &       -      &       \!$<0.1$\!      &       \!0.20\!     &    \!1292.9\!       &  -\\
    &\!\!MS-WACD\cite{Schmalenstroeer2017}\!\!        &       -      &       \!$<0.1$\!     &       \!0.25\!     &    \!1032.6\!       &  -\\
    &\!\!MS-DXCP\cite{Chinaev19}\!\!&       -      &       \!$<0.1$\!      &       \!$<0.1$\!     &    \!394.2\!       &  -\\ \cline{2-7}
    &prop. DXCP              &  \checkmark  &       \!$<0.1$\!      &       \!0.30\!     &     \!130.1\!       &  \!\bf 3.8\!\\
%    \hline
    \end{tabular}
    \caption{SRO estimators implemented in various architectures.}
    \vspace{-3mm}
    \label{tab:Evaluation}
\end{table}
\section{Conclusion}
\label{sec:conclusion}
\vspace*{-1ex}

For sampling-rate offset (SRO) assessment from discrete-time waveforms in asynchronous acoustic sensor networks, this paper picked up on the accurate time-domain double-cross-correlation processor (DXCP), which formerly has been available in merely an offline form. The contribution of the paper is then two-fold in that we a) have here delivered a translation of DXCP to an efficient STFT-based online form and b) have embedded this online DXCP into a closed-loop internal-model control (IMC) architecture with asyncronous resampling and step-SRO support. In comparison with several state-of-the-art methods, the proposed system unites the advantages of online processing, low complexity, and high SRO accuracy.

\bibliographystyle{IEEEbib}
\bibliography{refs}

\end{document}